# Communication conditions in virtual acoustic scenes in an underground station

Ľuboš Hládek
*Audio Information Processing*
*Technical University of Munich*
Munich, Germany
0000-0002-0870-7612

Stephan D. Ewert
*Medizinische Physik*
*CvO Universität Oldenburg*
Oldenburg, Germany
stephan.ewert@uol.de

Bernhard U. Seeber
*Audio Information Processing*
*Technical University of Munich*
Munich, Germany
seeber@tum.de

*Abstract*— Underground stations are a common communication situation in towns: we talk with friends or colleagues, listen to announcements or shop for titbits while background noise and reverberation are challenging communication. Here, we perform an acoustical analysis of two communication scenes in an underground station in Munich and test speech intelligibility. The acoustical conditions were measured in the station and are compared to simulations in the real-time Simulated Open Field Environment (rtSOFE). We compare binaural room impulse responses measured with an artificial head in the station to modeled impulse responses for free-field auralization via 60 loudspeakers in the rtSOFE. We used the image source method to model early reflections and a set of multi-microphone recordings to model late reverberation. The first communication scene consists of 12 equidistant (1.6 m) horizontally spaced source positions around a listener, simulating different direction-dependent spatial unmasking conditions. The second scene mimics an approaching speaker across six radially spaced source positions (from 1 m to 10 m) with varying direct sound level and thus direct-to-reverberant energy. The acoustic parameters of the underground station show a moderate amount of reverberation (T30 in octave bands was between 2.3 s and 0.6 s and early-decay times between 1.46 s and 0.46 s). The binaural and energetic parameters of the auralization were in a close match to the measurement. Measured speech reception thresholds were within the error of the speech test, letting us to conclude that the auralized simulation reproduces acoustic and perceptually relevant parameters for speech intelligibility with high accuracy.

*Keywords — sound field rendering, speech perception, room acoustics*

## I. INTRODUCTION

Creating an acoustically accurate reproduction of life-like acoustic scenes in a laboratory is a challenging task, but it is one of the key components for ecologically valid testing in environments for hearing research [1]. Traditionally, the protocols limit the interaction of the participants with the environment, which is reasonable in many psychoacoustic paradigms, but at the same time it constraints the generalization when it comes to studying communication in real-life situations. For instance, people experience less benefit form hearing aids in real-life situations than is expected from the laboratory studies [2], [3]. Increasing ecological validity by making acoustically comparable simulations of acoustic scenes has been considered in various previous works [4]–[13], many of which focused on speech perception, sound localization, audio quality and other aspects. These works analyze the standard acoustic parameters, psychoacoustic parameters, consider outputs of different psychoacoustic models, or assess these aspects directly in behavioral experiments. In the present work, we aim to recreate two communication scenes in a Munich underground station and evaluate the efficacy of our sound reproduction method following approaches established in room acoustics and psychoacoustics.

Binaural synthesis, synthesizing sounds with recorded binaural room impulse responses (BRIRs) and reproducing them over headphones, is a standard approach in psychoacoustic experiments. The recordings capture the acoustic information available at the ears and preserve binaural cues with a high degree of fidelity. However, the perception of sound over headphones can be qualitatively different from what people experience in reality. For instance, headphone reproduction can affect sound localization and externalization [14]. Highly realistic interactive headphone reproductions can be achieved, but this may require substantial effort, like the measurement of individualized head-related transfer functions and real-time processing. Such effort is complicating experiments with groups of elderly and hearing impaired people, who are often a target group in hearing research, but have limited availability [15].

In search for addressing the limitations of headphone reproduction, research has been devoted to develop loudspeaker-based spatial reproduction systems, e.g., [5], [9], [16], [17]. The loudspeaker reproduction can create an immersive percept of room acoustics: the subjects sit unencumbered by earphones in the free sound field and listen naturally with their own ears [4]. Controlled conditions can be achieved when the loudspeakers are distributed both horizontally and vertically in an anechoic chamber [16]. The limitations of this approach are usually the size of the acoustic sweet spot and the sparsity of the loudspeaker distribution, but also sound reflections off the equipment, which alter the sound field [18]. For loudspeaker reproduction, the sound scene can be acoustically modeled or recorded. An example of the modeling approach is the image source method [19], [20], which can accurately model early, specular reflections in rooms with various shapes. In many real rooms and source-listener configurations, the early part of room impulse response is dominated by specular reflections and these have high perceptual importance in terms of speech intelligibility [21], while sound location is usually dominated by the direct sound due the precedence effect [22], [23]. Recording the acoustics of sound scenes is possible with a multi-channel microphone array that can capture all reflections, early reflections and late reverberation. Reproduction can be achieved, for instance, by using the Ambisonics technique [24]. The microphone signals are transformed into the spherical harmonics domain and are subsequently decoded into loudspeaker signals. The method is limited by the physical properties of the recording device. Its dimensions define the spatial aliasing frequency. Despite that, Ambisonics, especially with higher orders, has been widely used and various methods have been proposed to minimize its reproduction artifacts [25].

In this work, we aim to acoustically recreate an underground station, investigate the acoustical properties of the simulation and compare speech intelligibility of the simulation with loudspeaker-based auralization against a baseline using binaural synthesis via headphones with impulse

Funded by the Deutsche Forschungsgemeinschaft (DFG, German Research Foundation) – Projektnummer 352015383 – SFB 1330, Project C5. rtSOFE development is supported by the Bernstein Center for Computational Neuroscience, BMBF 01 GQ 1004B.

responses recorded in the underground. The modeled impulse responses are used in free field loudspeaker reproduction. We assume that early reflections change rapidly in an acoustic scene with moving sources and listener but at the same time they could be efficiently computed with the image source method, even in real-time [16]. Late reverberation originating from many overlapping higher-order reflections can be statistically described and is considerably less affected by (small) movements of source and receiver than the early reflections. Therefore, late reverberation can be approximated by, e.g., fixed impulse responses obtained from fitted offline-models [26], or here by using a set of discrete multi-channel microphone array recordings [27].

## II. METHODS

### A. Acoustic scenes

We define two acoustic scenes at the platform of the Munich underground station Theresienstraße. A scene determines a set of source-listener configurations representing standard listening situations, which are shown in Fig. 1. The first scene represents a situation with horizontally placed sound sources at 1.6 m distance, a "Corona-safe" communication situation on the platform with a talker at an arbitrary horizontal direction. The second scene represents an approaching talker to the listener; the sound sources span radially in front of the listener from 10.1 m to 1.01 m such that the direct sound increases by 4 dB in each step (16 dB from position 17 to 13), which affects the direct-to-reverberant energy ratio (DRR) and listening in background noise.

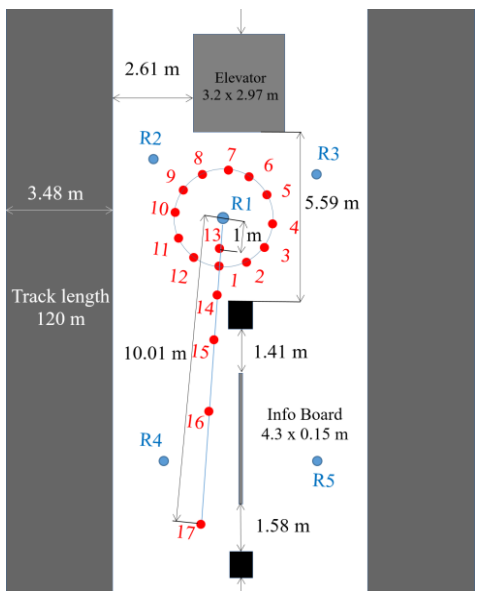

Fig. 1. Schematic distribution of sources (red, numbered) and receivers (R, blue) in the two acoustic scenes on the platform of the underground station. Positions 1-12 correspond to Scene 1, positions 13-17 to Scene 2. R1 corresponds to the listener position, R2-R5 are auxiliary measurement positions. All sources are facing the listener position R1. The listener was facing source position 1. Positions 1,3,13,14,16 were used in the speech perception experiment.

We performed acoustic measurements at selected positions defined in the scenes during the closing hours of the underground to achieve a high signal-to-noise ratio (SNR) [28]. To measure a set of impulse responses, 1-minute-long exponential sine sweeps excited a sound source (BM6A MKII, Dynaudio, Skanderborg, Denmark) that was placed at different positions, while an artificial head (HMS II.3-33, Head Acoustics, Herzogenrath, Germany) was at the listener position R1 (Fig. 1). Five these recordings were used for the headphone reproduction during the speech perception experiment. For a subset of source positions, we recorded impulse responses with an omni-directional microphone (MM210, Microtech Gefell, Gefell, Germany) and a multi-microphone array (EM32 Eigenmike, MH Acoustics, San Francisco, CA, USA) at the listener position. To estimate reverberation times according to [29], we measured an additional set of impulse responses with an omni-directional sound source (RAK4 - Typ DOD100B, Müller BBM, Munich, Germany) at two different positions and with the omni-directional microphones placed at five different positions (Fig. 1: R1-R5). Both sound sources were equalized in terms of frequency response prior to the recordings. The procedures and available data are detailed in van de Par et al. [28].

Table 1 summarizes the reverberation time ($T_{30}$) and early decay time (EDT) computed from the impulse response recorded with the omni-directional source and omni-directional microphone in the underground station. The analysis was conducted using the ITA Toolbox [30], data were averaged across five different measurements where distance from source to receiver was greater than 5 m. The data were analyzed in seven octave bands.

TABLE I. ACOUSTIC PARAMETERS

| | Frequency (Hz) | | | | | | |
|---|---|---|---|---|---|---|---|
| | *125* | *250* | *500* | *1000* | *2000* | *4000* | *8000* |
| $T_{30}$ (s) | 1.73 | 2.44 | 2.05 | 1.71 | 1.47 | 1.11 | 0.65 |
| EDT (s) | 1.02 | 1.46 | 1.42 | 1.19 | 1 | 0.8 | 0.46 |

### B. Sound reproduction

Environment (rtSOFE) [5], [16]. The rtSOFE was designed for realistic audio-visual reproduction and experimentation, however, in the paper we focus only on the acoustics, not the visual representation. The rtSOFE consists of 60 horizontally and vertically distributed loudspeakers (BM6A MKII, Dynaudio, Skanderborg, Denmark) inside an anechoic chamber. There are 36 horizontal loudspeakers (0° elevation at 1.4 m height) spaced evenly in 10 degrees and 24 elevated loudspeakers, twelve at -19° elevation and twelve at +32° elevation with 30° horizontal spacing. The loudspeakers are supported by a steel cube-shaped frame with dimensions of 4.8 m x 4.8 m x 3.9 m (l x w x h), and they are driven by DA converters RME 32DA (Audio AG, Haimhausen, Germany) connected to a multi-channel sound card RME HDSPe (Audio AG, Haimhausen, Germany). The loudspeakers were equalized in terms of time, amplitude, and phase, in the frequency range 90 Hz – 18 kHz each with an FIR filter with 1024 taps with respect to the center of the loudspeaker array. A binaural presentation system consists of headphones (HD 650, Sennheiser, Wedemark, Germany) connected to a custom made headphone amplifier with 24 bit DA converter. The headphones were equalized with a 512-tap FIR filter in the range of 90 Hz – 14 kHz in terms of amplitude. The calibration procedure was conducted using the artificial head with a real ear simulator. The sampling frequency was 44.1 kHz.

### C. Acoustic modeling

We modeled the scenes following the assumption that the room impulse response consists of specular early reflections and (diffuse) late reverberation. Early reflections were modeled with the image source method, while the late reverberation is homogeneous near the recording site, and therefore we modeled the reverberant tail using a set of recorded spatial impulse responses.

**Early reflections** - we created a spatially accurate model based on a 3D laser scan of the underground station and modeled early reflections using the image source method as implemented in the rtSOFE (v4) [5], [16]. The model considered the directivity of the sound source (separately measured in the anechoic chamber) and environmental properties during the recordings (temperature, humidity, speed of sound). Each reflection (impulse) was modeled with a FIR filter with order 380 (256 was used for one position). Individual reflections were then mapped to 36 horizontal loudspeakers using 17th-order Ambisonics (horizontally) with the sampling decoder [31]. The reflective properties of the walls in the model were obtained from the text book values that approximated the properties of materials in the underground station.

**Late reflections** – we obtained a collection of spatialized impulse responses in four discrete positions (1,3,11,16) using a multi-microphone array Eigenmike EM32. These impulse responses were used to simulate late reverberation. First, we created a representation of the recorded impulse response signal in the spherical harmonics domain, which we mapped to the thirty-six horizontal and vertical loudspeakers of rtSOFE. We used Array2SH (ver 1.6.6) [32] and AmbiDEC (ver 1.6.5) Sparta plugins [33] in Reaper (v6.23, Cockos Inc., Rosendale, NY, USA) to create the loudspeaker signals, which were than transformed to spatialized impulse responses. Second, we created a filter that equalized the overall frequency response of the multi-channel recording to the frequency response of the omni-directional microphone recording at the identical positions. The frequency response of the resulting FIR filter of order of 512 is shown in Fig. 2.

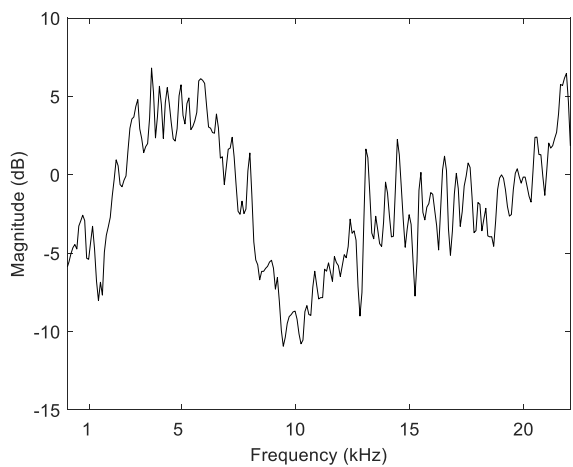

Fig. 2. Multi-channel recording equalization filter.

**Early-Late separation time (E-LST)** – The early and the late reflections were combined into one spatialized impulse response for loudspeaker reproduction in rtSOFE. The E-LST refers to a time that separates the early and late part of the impulse response [21]. Similar nomenclature is used with the room acoustic parameter definition (see [29]), which represents ratio of energy of the early part and total energy. In this work, we aimed to identify the separation time such that the modeled impulse response could truthfully approximate the measured impulse response in terms of standard acoustic parameters important for speech communication: reverberation times ($T_{20}$, $T_{30}$), clarity indices ($C_{50}$, $C_{80}$), and early decay time (EDT).

To analyze the effect of E-LST on the parameters, we created a modified modeled impulse response, such that the late part consisted of the impulse response recorded with an omni-directional microphone and the early part consisted of the modeled specular reflections (from image source method). In this analysis, we considered the directivity of the omni-directional microphone. The mixing of the two parts of the impulse response, the early reflections and late reverberation, was done with short onset-offset linear ramps with duration of 2 ms with a cross-fade. Then, by setting the E-LST to 0 s, one has the reference omnidirectional recording and by increasing the parameter, we see the effect of modeling on the acoustic parameters. The aim of the analysis is to identify an appropriate E-LST value, such that the resulting impulse response preserves the acoustic properties of the environment while early reflections permit dynamic binaural rendering.

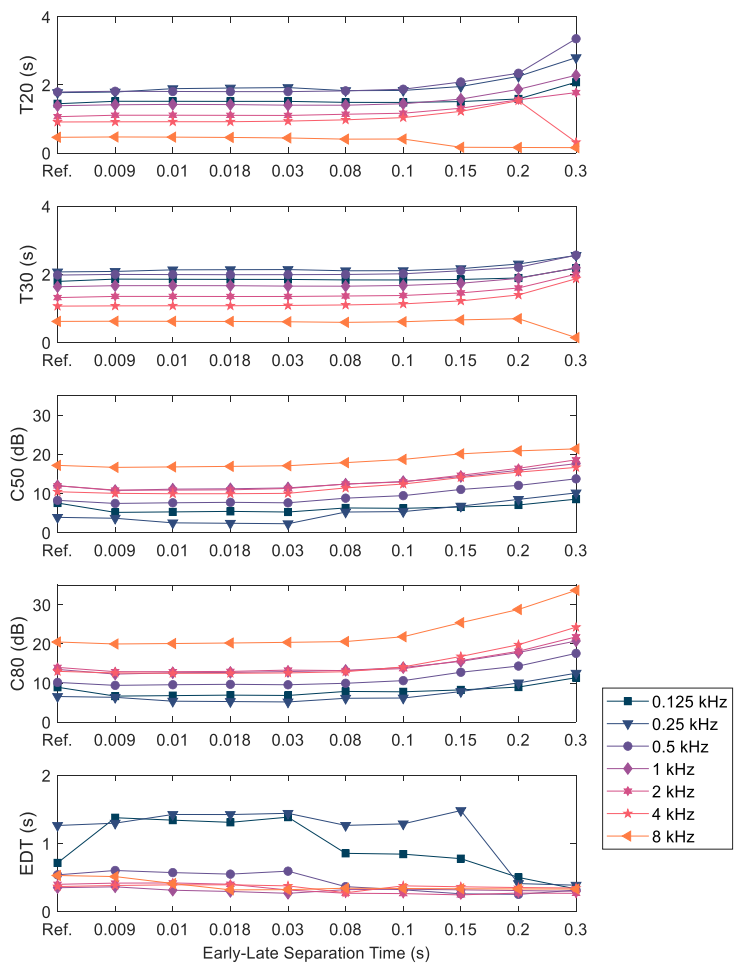

Fig. 3. Effect of Early-Late Separation time on acoustic parameters reverberation time $T_{30}$, $T_{20}$, clarity indices $C_{50}$, $C_{80}$, and early decay time (EDT) in different octave bands. The reference values (Ref. on y-axis) are obtained from an impulse response recorded with an omni-directional microphone at source position 1.

The top two panels of Fig. 3 show the effect of the E-LST on the reverberation times T20 and T30. The data show very little variation from the reference value for the values of E-LST up to 100 ms. Further increase of E-LST leads to a variation of the values across frequency bands. The middle two panels of Fig. 3 analyze the effect of E-LST on the clarity indices. Here, we see a slight variation of these values in some frequency bands for E-LST up to 100 ms but the variations are small, usually below 2 dB from the reference. E-LST values higher than 100 ms show an increase for $C_{80}$ across frequency bands. The bottom row shows the values of EDT. The EDT show energetic contribution of the first reflections (up to -10dB of the energy decay curve), the values of EDT which are below reverberation time suggest a higher energetic contribution of reflections than the exponential decay as measured by reverberation times. This may have a positive effect on speech reception thresholds (SRT), and therefore it is an important parameter in acoustic design. The values of EDT show only little variations up to 150 ms except the frequency band with the 125 Hz center frequency, which may relate to the frequency limit of the loudspeaker used in the measurement. However, the values at 100 ms and 150 ms show a close match with the reference. Taken together, early reflections simulated as specular reflections with the image source method capture the acoustic parameters of the underground station well for the first 100-150 ms of the impulse response, while for later reflections a non-specular reflection model could provide a better approximation to the measured impulse response. Hence, the present study uses a hybrid impulse response model with simulated early reflections up to 100 ms and measured late reverberation.

*D. Speech perception test*

To see how well the modeled spatial impulse response recreates the acoustics of the scenes for binaural speech perception, we compared SRT in the modeled scene presented in the rtSOFE over loudspeakers to those of the recorded binaural scene presented over headphones. We created a set of standard source-listener configurations for both scenes to see the effect of distance, reverberation and spatial separation of the target and the interferer. The noise interferer was kept at a constant position, source position 1 (in front of the listener, zero degrees, "N0"), while the target speech was presented at different positions to simulate the changing position of a talker (see below). In Scene 1, we choose positions 1 and 3 to see the effect of horizontal separation. We expected a decrease in SRT at position 3 relative to position 1 due to spatial release from masking (SRM) that is limited by the reverberation of the interferer. In this situation, the energetic effects at the better ear and the binaural effects will be responsible for the SRM. In scene 2, we chose positions 13, 14, 16 (and 1), where the effect of spatial separation is mainly driven by the changing distance and the binaural and energetic effects of reverberation.

*1) Participants*

Six young volunteers (2 females), native German speakers, took part in the speech perception experiment. Their pure-tone thresholds at standard audiological frequencies (250 Hz – 8 kHz) were assessed with a calibrated audiometer (MADSEN Astera², type 1066, Natus Medical Denmark Ap, Denmark) and were below 20 dB HL. The ethics committee of the Technical University of Munich approved the methodology of the study (ref: 65/18S).

*2) Stimuli and Procedures*

The participants performed a customized version of the Oldenburg Sentence test (OLSA) [34]. The OLSA is a standardized matrix speech perception test, where five-word sentences are formed from a closed set of ten options for each word. The difference to the original OLSA was that participants had a GUI with all possible options available at all times. The stimuli consisted of sentence lists from the OLSA CD. Individual sentences were presented at varying sound levels to estimate the SRT. The SRT was defined as 50% correct of the number of words of a sentence. The level of speech was intended to be clearly audible at the beginning of each track; the initial values were determined during the pilot stage and were set to +6 dB above the expected threshold. The speech position varied for different tracks, and it could be one of five source positions: 1 (0° (front), 1.6 m distance), 3 (60° left, 1.6 m distance), 13 (0° (front), 1.01 m distance), 14 (0° (front), 2.53 m distance) or 16 (0° (front), 6.37 m distance). The interfering noise was a speech-shape modulated noise, 'Fastl noise', [35], [36] presented at 65 dB SPL (reference level @ 1 m distance); it was always presented from source position 1 (front 1.6 m distance). The stimuli were spatialized by convolving the sound with recorded or modeled spatial impulse response at the corresponding source positions. The stimuli spatialized with the recorded impulse responses were presented over the *headphones (HP)*, the stimuli spatialized in the modelled scene were presented over the *loudspeakers (LS)*.

The experiment consisted of 10 blocks, each collecting data in one adaptive track. During one adaptive track, one sentence list was used and the SRT was estimated from the last 20 sentences. Each of the five positions was run in two modes of presentations. Each block had a random order of sentences and fixed mode of presentation. However, the mode of presentation alternated between blocks. Participants were assigned to one of two groups, such that one of the groups started with the LS presentation and the other with the HP presentation to eliminate any order effects. The order of blocks and selection of lists was randomized for each participant in a way that each participant was randomly assigned five different lists, which were fixed with the positions (for one position the same list was used for HP and LS). Two identical lists were never presented successively.

During the experiment, participants were sitting on a swivel chair in the middle of SOFE and were holding a touch-screen tablet with the custom GUI, which displayed all possible word options of the OLSA. At the beginning of each block, the GUI informed them about the mode of presentation and they had to press the 'Start' button once they were ready to start. On each trial, participants heard a sentence, provided an answer into the GUI, and pressed the 'Next' button after completing all options. After completing a block, participants were encouraged to take a short break. Participants were instructed to keep their head straight during LS presentation. Prior to the experiment, all participants underwent four training blocks with the collocated speech and interferer (position 1), two with the LS and two with the HP mode of presentation. The sentence lists used during the training were not used during the main experiment. The whole session lasted about 2.5 hours.

*E. Analysis*

The SRTs for each of the conditions were statistically analyzed using a paired t-test. The analysis scripts were written in MATLAB (ver. 9.9, Mathworks, Natick, MA, USA). Further, we evaluated monaural and binaural parameters: speech-weighted direct-to-reverberant energy ratio (DRR), interaural time difference (ITD), interaural level difference (ILD) and interaural coherence in different positions of the modelled and recorded impulse responses. The parameters of the modelled scene (LS) are evaluated against the reference values obtained from the in-situ binaural recordings (HP), and compared to the anechoic versions of the impulse responses. The parameters for the HP mode were obtained directly from the recorded binaural impulse responses, the parameters for the LS mode were obtained by convolving the loudspeaker signals with head-related impulse responses measured with the artificial head for each SOFE loudspeaker and summing them across the loudspeakers.

We defined the anechoic signal as the signal with only the direct sound and no reflections. The time of the first reflection was identified visually in each impulse response. This was used also for computing the speech-weighted DRR. The ITD was computed as the lag of the peak of the interaural cross-correlation function (with a limit of ±1 ms) of the ear signals low-pass filtered at 1300 Hz. The interaural coherence was computed as the peak value of the normalized interaural cross-correlation function of the low-pass filtered signals. The ILD was computed as an energy ratio of the left and the right ears of the high-passed signal at 1300 Hz.

III. RESULTS

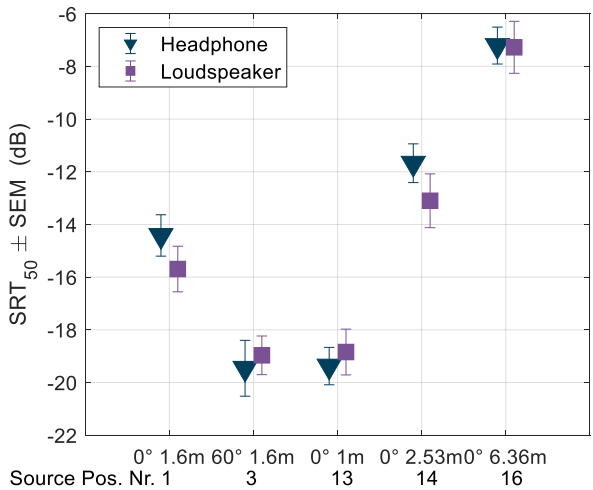

Fig. 4. Speech reception thresholds for loudspeaker (LS, ■) and headphone (HP, ▼) presentation. The x-axis shows different sound source positions. Scene 1 – positions 1 and 3, Scene 2 – positions 13, 14 and 16. The caption denotes angle and distance relative to the listener position. Noise was always at position 1.

The SRT at the position with the co-located target and interferer (source position 1) is at -14.4 dB for HP (Fig. 4). Although it is a highly negative SRT, such values could be explained by the nature of the interferer and the use of a closed-set test. In Scene 1, we tested an additional position, source position 3. The comparison relative to the position 1 yields the SRM, which is 5 dB (for HP). The values of the LS and HP modes are within the across-subject variance (two-tailed paired t-test: p>0.05 for all positions).

For Scene 2, all positions are on a straight radial line. The distance-related change of the sound level of the direct sound for positions 13, 14, and 16 are 4 dB, -4dB, -12 dB. The SRM for these positions equals to 5 dB, -2.7dB, -7.2 dB, respectively. The change of SRT in position 13 re. position 1 could be explained by the distance related change. The two positions behind the interferer (positions 14 and 16) overestimate (people perform better) what is expected based on the distance-related level change. This can be mainly accounted to the effects of reverberation on the speech signal, which also captures the directivity of the sound source.

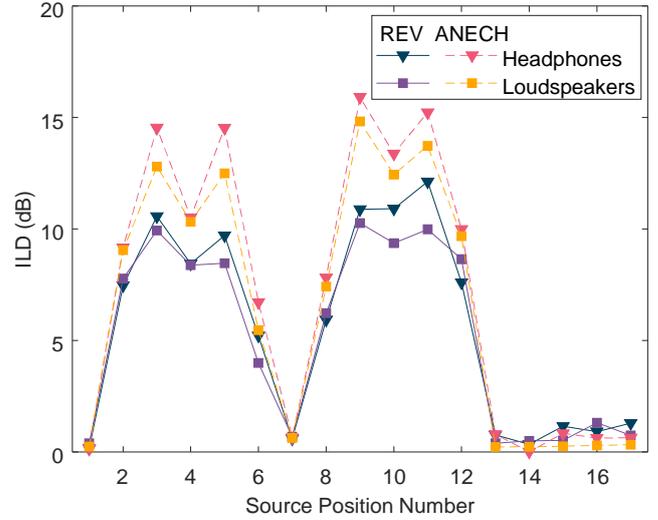

Fig. 5. Interaural level differences for different positions in Scene 1 (source positions 1-12) and Scene 2 (source positions 13-17). Solid lines show the data for reverberant space using the recorded baseline (HP) and the modelled loudspeaker auralization (LS). Dashed lines show data from the direct sound only, i.e. anechoic impulse responses.

The data of ILDs (Fig. 5) show an expected increasing pattern with laterality for Scene 1, the positions spanning radially (13-17) show only a subtle change of ILD possibly due to early reflections. We see only a small difference between the HP and the LS and only for a few positions (5,10,11), which could be related to Ambisonics reproduction artifacts at high frequencies with LS reproduction. Further, the data show that the reverberation decreases the ILDs by about 5 dB for the lateral positions in Scene 1 and increases ILD by no more than 2 dB in radial positions in Scene 2.

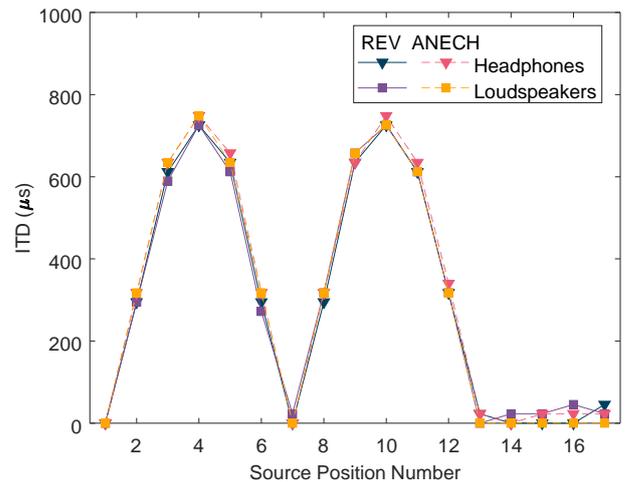

Fig. 6. Interaural time differences. The layout follows Fig. 5.

The data shown an expected pattern for the lateral and radial targets, there is only a small difference between LS and HP. The ITD values are mainly affected by the direct sound. The small deviations from the anechoic values could be related to early reflections.

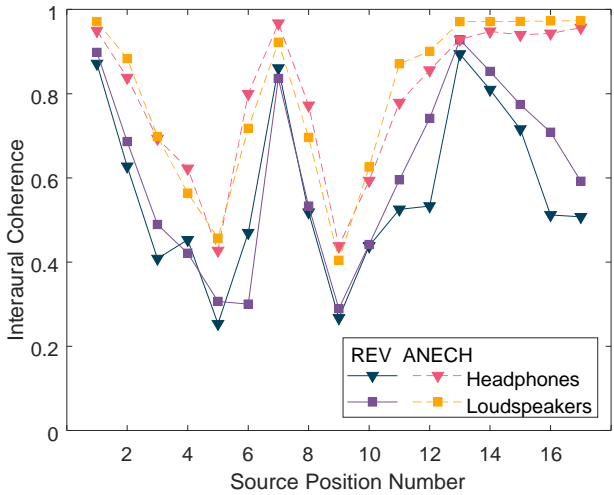

Fig. 7. Interaural coherence. The values on the graphs show the maxima of the interaural cross-correlation (IACC) function. They layout follows Fig. 5.

The interaural coherence strongly varies with the laterality of the sound source, source distance and reverberation. In Scene 1, the values vary between 0.87 at the front and 0.23 at the side of the listener. In Scene 2, the values vary between 0.89 (positon 13) and 0.5 (position 17). The HP and LS modes agree fairly well although coherence is slightly higher in some position for LS compared to HP presentation, especially in Scene 2, but all within the just noticeable difference for the IACC.

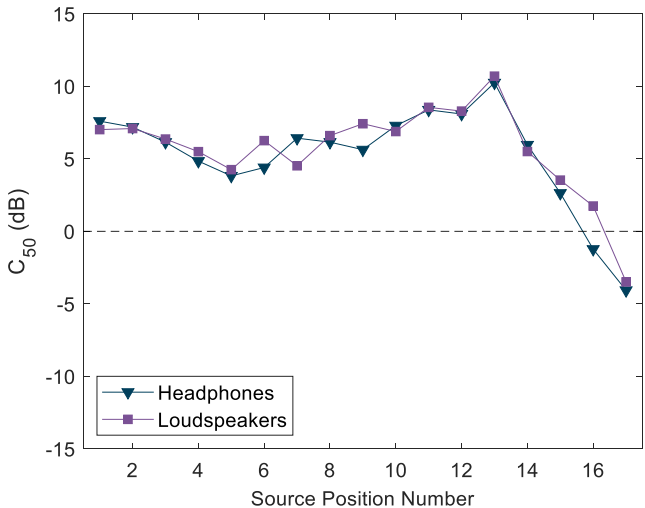

Fig. 8. Speech-weighted clarity index (C50) computed as the mean based on the left ear and the right ear impulse responses. The x-axis shows different measurement positions.

Fig. 8 shows that the clarity index C50 has highly positive values in rage 3.8 dB to 8.2 dB for Scene 1 (distance is fixed to 1.6 m) and the values vary between 10 dB to -4dB for Scene 2 (egocentric distances vary between 1.01 m and 10.1 m). There is a close agreement between the LS and HP mode of presentation with only slight variations for some source positions.

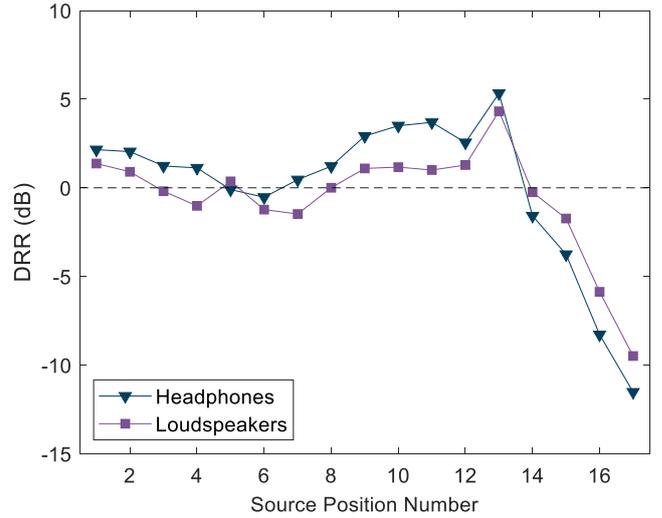

Fig. 9. Speech-weighted DRR computed as the mean of DRRs for the left ear and the right ear impulse responses. The x-axis shows different measurement positions.

Fig. 9 shows that speech-weighted DRR for binaural signals. For the Scene 1 (1-12), the values were mostly positive and varied between -0.5 dB and 3.6 dB (HP mode), depending on the laterality (egocentric distance is fixed to 1.6 m). The values for Scene 2 (13-17) varied between 5.3 dB and -11.5 dB (HP mode), showing the effect of egocentric distance (from 1.01 m to 10.1 m). For Scene 1, the LS mode slightly underestimates the values obtained from the recorded impulse responses (1dB - 2dB), and over estimates the values for the Scene 2. The difference could relate to the Ambisonics reproduction of the late reverberation and small imprecisions in the image source model.

## IV. DISCUSSION

In this work, we recorded and recreated the acoustics of a Munich underground station where we defined two scenes with different source-receiver location combinations. Scene 1 distributes sources circularly around the listener, in Scene 2 the sources are at different distances on a line in front of the listener. The nearby positions of both scenes have a moderate amount of reverberation, the early decay times were lower than $T_{30}$ values, indicating that early reflections have strong energy. We investigated speech perception in a set of standard source-listener configurations and compared two modes of reproductions, a headphone baseline (HP) and a simulation using free-field loudspeaker auralization (LS). The HP condition was based on binaural recordings with an artificial head, the LS condition on modeling of the early (image source model) and the late reverberation (spatialized recording). In an analysis, we defined the transition time between early, specular reflections and reverberation (E-LST) to 100 ms, which seems to preserve the acoustic properties of the environment.

Measured speech reception thresholds for the recorded and modeled acoustics were within the across-subject variance. This suggests that the LS mode provides the acoustic features in a similar way as the recorded reference. We confirm this by an analysis of the interaural and energetic cues. When the speech target and noise interferer were spatially separated in horizontal plane, Scene 1, we observed a high degree of correspondence between the two modes of reproduction in terms of ITDs, ILDs and interaural coherence.

In Scene 2, the SRTs varied with distance of the target speech source. Here the presence of the reverberation had a positive effect on SRTs, leading to less reduction with distance as expected from the direct sound alone. This could relate to the energetic effects of the early reflections on the target speech signal. Analysis of DRR for positions of 13-17 shows a steep decrease of about 16 dB, suggesting that the change of DRR is driven mainly by the energetic change of the direct sound, as would be expected in an environment with approximately constant reverberant energy. On the other hand, $C_{50}$ shows shallower decrease, suggesting that early reflections are likely to energetically contribute to the speech signal, which closely relates to observations of SRTs. The effect of late reverberation on speech is often considered as detrimental, which is a standard view in terms of acoustics and speech intelligibility modelling [21]. However, for specific situations as the one used here, diffuse reverberation can aid unmasking for improved speech intelligibility [37] and tone detection [38]. This points to a more complex spatio-temporal effects of reverberation than the standard model based on discrete division of the impulse response to positive early reflections and detrimental late reverberation.

The analysis of 17 positions at different azimuth angles and distances suggests the present results generalize also to untested positions within the same area, and we can expect that our modeling approach would recreate acoustical conditions that provide reasonably accurate cues for speech perception.

In this work, we demonstrated that the loudspeaker-based auralization in rtSOFE recreates the speech-perception relevant acoustical features of the underground station. Since the method was based on modeling, we can further use this approach for real-time reproduction of the scene and modelling of dynamic scenarios with moving sound sources. In the next step, we aim to evaluate sound localization and sound quality properties and recreate the scene for interactive communication experiments including visual representation.

## V. ACKNOWLEDGMENT

We thank prof. Steven van de Par from Oldenburg University for providing the multi-microphone array.